\documentclass[english,aps,prl, twocolumn,superscriptaddress]{revtex4-1}
\usepackage[utf8]{inputenc}
\usepackage{amsmath}
\usepackage{amssymb}
\usepackage{bm}
\usepackage{graphicx}
\usepackage{braket}
\usepackage{babel}
\usepackage{mathtools}
\usepackage{xspace}
\usepackage{xcolor}
\usepackage{physics}

\usepackage[backref=none,
bookmarksnumbered=true,
bookmarks=true,
bookmarksopen=true,
colorlinks=true,
citecolor=blue,
linkcolor=blue,
anchorcolor=green,
urlcolor=blue,unicode=false]{hyperref}

\newcommand{\even}{^\mathrm{e}}
\newcommand{\odd}{^\mathrm{o}}

\begin{document}

\title{Robust spectral $\pi$ pairing in the random-field Floquet quantum Ising model}

\author{Harald Schmid}
\affiliation{\mbox{Dahlem Center for Complex Quantum Systems, Freie Universit\"at Berlin, 14195 Berlin, Germany}}

\author{Alexander-Georg Penner}
\affiliation{\mbox{Dahlem Center for Complex Quantum Systems, Freie Universit\"at Berlin, 14195 Berlin, Germany}}

\author{Kang Yang}
\affiliation{\mbox{Dahlem Center for Complex Quantum Systems, Freie Universit\"at Berlin, 14195 Berlin, Germany}}

\author{Leonid Glazman}
\affiliation{\mbox{Department of Physics, Yale University, New Haven, Connecticut 06520, USA}}

\author{Felix von Oppen}
\affiliation{\mbox{Dahlem Center for Complex Quantum Systems, Freie Universit\"at Berlin, 14195 Berlin, Germany}}

%
%
%
\begin{abstract}
Motivated by an experiment on a superconducting quantum processor [Mi et al., Science \textbf{378}, 785 (2022)], we study  level pairings in the many-body spectrum of the random-field Floquet quantum Ising model. The pairings derive from Majorana zero and $\pi$ modes when  writing the spin model in Jordan-Wigner fermions. Both splittings have lognormal distributions with random transverse fields. In contrast, random longitudinal fields affect the zero and $\pi$ splittings in drastically different ways. While zero pairings are rapidly lifted, the $\pi$ pairings are remarkably robust, or even  strengthened, up to vastly larger disorder strengths. We explain our results within a self-consistent Floquet perturbation theory and study implications for boundary spin correlations. The robustness of $\pi$ pairings against longitudinal disorder may be useful for quantum information processing.  
\end{abstract}

\maketitle 

{\em Introduction.---}The quantum Ising model \cite{Pfeuty1970} appears at the crossroads of many current developments in  condensed matter physics and quantum information. It is paradigmatic for symmetry breaking quantum phase transitions in its spin incarnation \cite{Sachdev2011}, for topological quantum phase transitions in its fermionized version \cite{Kitaev_2001}, for lattice gauge theory as well as topological quantum error correcting codes in its dualized form \cite{Kogut1979,Kitaev1997},
and for time crystals as a Floquet model \cite{Khemani2016,Else2016}. A recent experiment on a superconducting quantum processor \cite{Mi2022} reveals that temporal spin correlations of the one-dimensional Floquet quantum Ising model can be remarkably robust against certain types of disorder. 

\begin{figure*}[t]
    \centering
\includegraphics[width=\textwidth]{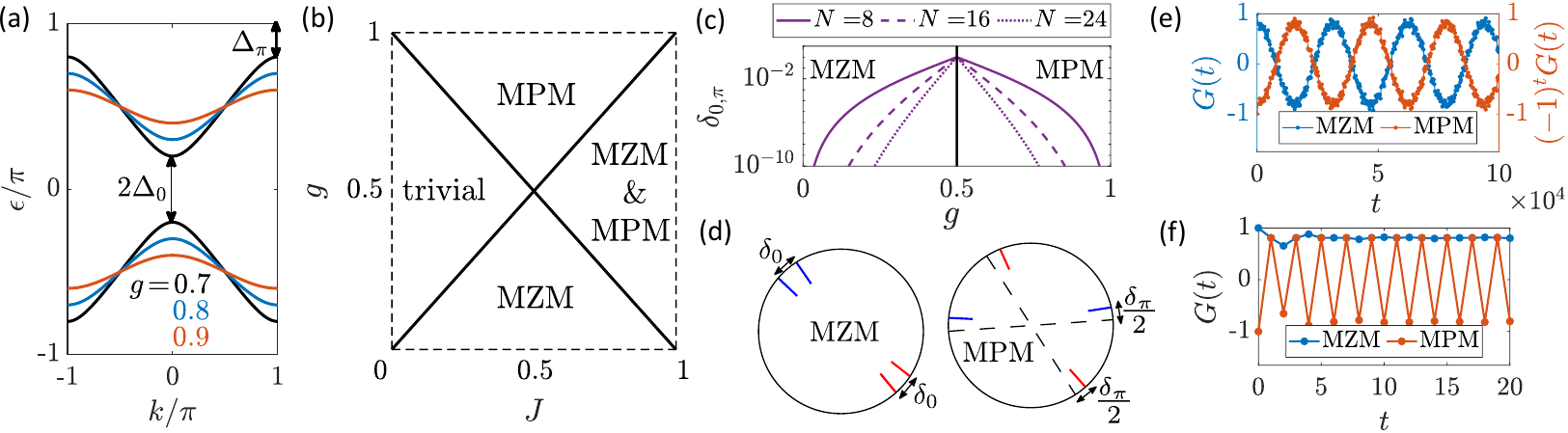}
\caption{Clean quantum Ising chain. (a) Single-particle Floquet spectrum of a periodic chain vs.\ wave vector $k$ for various transverse fields $g$. (b) Phase diagram with phases labeled by the Majorana modes present in the corresponding Kitaev chain. (c) Hybridization splitting $\delta_{0,\pi}$ vs.\ $g$ of Majorana modes in finite chains of various lengths $N$ ($J=0.5$), showing the symmetry between MZM and MPM phases. (d) Sketch of the pairing of many-body eigenphases in the MZM and MPM phases. Both pairings coexist in the MZM \& MPM phase. (e), (f) Spin-spin correlation function $G(t)$ for (e) long (note the factor $(-1)^t$ for the MPM phase) and (f) short times. For long times, $G(t)$ oscillates with period  $2\pi/\delta_{0,\pi}$, superimposed on rapid period-two oscillations in the MPM phase. Parameters: (a) $J=0.5$, (e), (f) $N=8$, $(g,J)=(0.2,0.5)$ (MZM), $(g,J)=(0.8,0.5)$ (MPM).}
\label{fig1}
\end{figure*}

In one dimension and in the absence of disorder, the Floquet quantum Ising model is defined through the Floquet operator 
\begin{equation}
\label{eq:U_F0}
    U_{F,0} = e^{{i\pi g\over 2}\sum_{j=1}^N X_j}
 e^{{i\pi J \over 2}
    \sum^{N-1}_{j=1}
     Z_j Z_{j+1}},
\end{equation}
which describes the stroboscopic time evolution of an initial state $\ket{\psi (0)}$ of $N$ qubits through $\ket{\psi (t)} = (U_{F,0})^t \ket{\psi (0)}$ with $t \in \mathbb{N}$. The Floquet operator $U_{F,0}$ can be implemented on a superconducting quantum processor through a set of single- and two-qubit gates. The two-qubit gates effect the Ising exchange coupling involving the Pauli-$Z$ operators of the qubits, while the single-qubit gates realize the transverse field in terms of the Pauli-$X$ operators. The model exhibits four topologically distinct phases as a function of the transverse field $g$ and the exchange coupling $J$ \cite{Dutta2013,Bauer2019,Lerose2021,SI}. This can be seen by diagonalizing $U_{F,0}$ by a Jordan-Wigner mapping to the Floquet Kitaev chain, a free-fermion model. For periodic boundary conditions, the single-particle eigenstates of the associated Bogoliubov-de Gennes Floquet operator can be labeled by momentum. The corresponding spectrum of eigenphases $\epsilon\in[-\pi,\pi]$ is shown in Fig.\ \ref{fig1}(a). One finds two gaps, one  around $\epsilon=0$ and another around $\epsilon=\pm \pi$, which can both be trivial or topological. This results in the four possible phases displayed in the phase diagram in Fig.\ \ref{fig1}(b) \cite{Dutta2013,Bauer2019}. 

In an open chain, the two types of topological gaps are signaled by a pair of Majorana zero modes (MZMs) or Majorana $\pi$ modes (MPMs), respectively \cite{Jiang2011}. These modes appear in the middle of the corresponding gap and exhibit a hybridization splitting away from $\epsilon=0$ (MZMs) or  $\epsilon=\pm \pi$ (MPMs) by an amount which is exponentially small in the length $N$ of the chain, see Fig.\ \ref{fig1}(c). In the presence of the Majorana modes, the corresponding many-body Floquet eigenstates of $U_{F,0}$ have eigenphases that come in pairs. Apart from hybridization splittings, the paired eigenphases are degenerate (MZMs) or shifted relative to each other by $\pi$ (MPMs), see Fig.\ \ref{fig1}(d). This is a particular instance of the wider phenomenon of strong modes in interacting and kicked spin models \cite{Fendley2016,Else2017,Kemp2017,Mitra2019,Yeh2023}.

Motivated by  experiment  \cite{Mi2022}, we study the effects of quenched random fields on these pairings of eigenphases as well as the ramifications for temporal spin-spin correlation functions. This is of considerable interest for two reasons. First, inaccuracies in implementing the gate operations naturally introduce a certain degree of randomness, making robustness against disorder an important issue in experiment and applications. Second, in the context of studying strong modes disorder raises important theoretical questions, especially because a random longitudinal field involving the Pauli-$Z$ operators breaks the protecting spin-flip symmetry of the clean quantum Ising model. Remarkably, we find that longitudinal disorder can even strengthen the spectral $\pi$ pairing, a result which extends beyond the robustness observed in experiment \cite{Mi2022}. We uncover dramatic differences between MZMs and MPMs, which may make the latter particularly interesting in the context of quantum information processing.

{\em  Random transverse field.---}We begin by studying random transverse fields and consider the Floquet operator $U_F=U_g U_{F,0}$ with $U_g=\exp\{\frac{i\pi }{2}\sum^N_{j=1}g_j X_j\}$. The random fields $g_j$ are drawn from independent box distributions, $g_j \in [-dg,dg]$. Unlike in related models of Floquet time crystals \cite{Khemani2016,Else2016,Keyserlingk2016a,Yao2017,Mi2021,Krzysztof2017,Khemani2019,Else2020,Zaletel2023} we consider a fixed $J$. Given that $U_g$ describes a field that is random in space but independent of time $t$, the disordered model remains Floquet and is characterized by a many-body spectrum of $2^N$ eigenphases $E_n$ on the unit circle, $U_F\ket{n}=e^{-iE_n}\ket{n}$.

In the presence of the random transverse field, one can still find a set of $N$ fermionic Bogoliubov operators $\gamma_\alpha$ satisfying 
\begin{equation}
    U_{F}^\dagger \gamma_\alpha U_{F}^{\phantom{\dagger}} = e^{-i\epsilon_\alpha} \gamma_\alpha. 
\end{equation}
This can, e.g., be done by expressing the spins in Jordan-Wigner fermions and a subsequent Bogoliubov transformation \cite{Sachdev2011,Mi2022} (as reviewed in the Supplementary Information \cite{SI}). Then, the $2^N$ many-body eigenphases $E_{n}= \sum_\alpha n_\alpha \epsilon_\alpha$ of  $U_{F}$ can be decomposed into the $N$ single-particle eigenphases $\epsilon_\alpha$. Here, the $n_\alpha\in\{0,1\}$ denote occupations $\gamma_\alpha^\dagger\gamma_\alpha^{\phantom{\dagger}}$ of the Bogoliubov fermions. Both $E_n$ and $\epsilon_\alpha$ are defined modulo $2\pi$. The above-mentioned zero ($\pi$) pairing of many-body states follows from the existence of a pair of MZMs (MPMs), which combine into a Bogoliubov fermion $\gamma_0$ ($\gamma_\pi$). The corresponding eigenphase $\epsilon_0$ ($\epsilon_\pi$) differs from  zero ($\pi$) by an amount $\delta_0$ ($\delta_\pi$) that is  exponentially small in the length of the chain. This leads to deviations from the perfect zero ($\pi$) pairing of many-body states by $\delta_0$ ($\delta_\pi$), which are identical for all pairs of the many-body spectrum. 

\begin{figure*}[t]
\centering
\includegraphics[width=\textwidth]{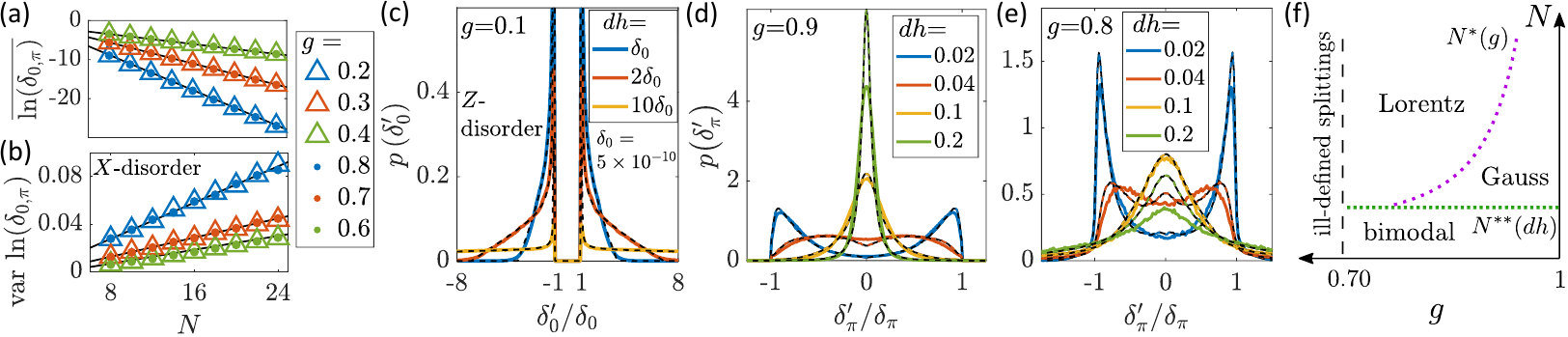}
    \caption{(a),(b) Random transverse fields: (a) Average and (b) variance of $\ln \delta_{0,\pi}$ vs.\ chain length $N$ for both MZMs ($g<1/2$; triangles) and MPMs ($g>1/2$; dots). Numerical results (symbols) are in excellent agreement with analytical expressions (full lines) \protect\cite{SI}. (c)-(e) Random longitudinal fields: Splitting distributions for various disorder strengths in (c) MZM  and (d),(e) MPM phase. In (c), numerical results (full lines) are well reproduced by an analytical two-level approximation (dashed lines). In (d),(e), numerical results (full lines) can be interpreted in terms of second-order Floquet perturbation theory (dashed lines). (f) ``Phase diagram" of the splitting distribution (MPM phase) in the $N-g$-plane for fixed $dh$.
    Parameters:  $J=0.5$, (a),(b) $dg=0.02$, $\mathcal{N}=10^4$ realizations, (c) $\delta_0=5\times 10^{-10}$, (d) $\delta_\pi=5\times 10^{-10}$, (e) $\delta_\pi=2\times 10^{-6}$, (c)-(e) $N=12$, $\mathcal{N}=10^3$.
    }
\label{fig2}
\end{figure*}

Random transverse fields induce a broad distribution of the splittings $\delta_0$ and $\delta_\pi$ across the disorder ensemble, which we find to be lognormal. Just as for the splittings in the clean model [see Fig.\ \ref{fig1}(c)], we find that the lognormal distribution for $\delta_0$ at $g$ is identical to the distribution of $\delta_\pi$ 
at $g\to 1-g$. This is illustrated in Fig.\ \ref{fig2}(a,b), which shows the average and variance of $\ln\delta_{0,\pi}$ as a function of $N$ for corresponding locations in the MZM and MPM phases. The linear dependence on $N$ reflects the exponential dependence of the hybridization splitting. The supplement \cite{SI} gives analytical expressions drawing on the related Hamiltonian problem \cite{Brouwer2011}.

{\em  Random longitudinal field.---}We now turn to the case of a random longitudinal field as described by the Floquet operator $U_F=U_h U_{F,0}$, where $U_h=\exp\{\frac{i\pi }{2}\sum^N_{j=1}h_j Z_j\}$ and  the random fields $h_j$ are drawn from independent box distributions, $h_j \in [-dh,dh]$. Longitudinal fields differ fundamentally from transverse fields in two ways. First, longitudinal fields do not conserve the spin-flip symmetry $P=\prod_j X_j$ of $U_{F,0}$, which maps to conservation of fermion parity $P=\prod_\alpha(1-2\gamma^\dagger_\alpha\gamma^{\phantom{\dagger}}_\alpha)$ in the Floquet Kitaev chain. As a result, longitudinal fields directly couple the two many-body states within a pair. Second, the fermionic representations of the Pauli-$Z$ operators involve string operators, turning the Floquet quantum Ising model with longitudinal disorder into an interacting fermion problem. 

Our numerics show that in stark contrast to transverse fields, random longitudinal fields affect the zero and $\pi$ splittings in dramatically different ways. In the MZM phase, even tiny random longitudinal fields of the order of $\delta_0$ enhance the splittings as shown in Fig.\ \ref{fig2}(c). We also find that the splittings remain approximately uniform across the many-body spectrum. In contrast, in the MPM phase, random longitudinal fields of order $\delta_\pi$ have essentially no effect. Even fields approaching order unity barely enlarge the splittings $\delta_\pi$. The splittings are strictly reduced when $g$ is sufficiently close to unity [Fig.\ \ref{fig2}(d)] and remain concentrated around zero when $g$ is further from unity [Fig.\ \ref{fig2}(e)]. The splittings vary across the many-body spectrum and are self gaveraging \cite{SI}. 

The remarkable robustness of MPMs against a random longitudinal field (as well as the sensitivity of MZMs) can be understood within a low-order stroboscopic Floquet perturbation theory for $U_F=e^{-i V}U_{F,0}$. Expanding the eigenphases of $U_F$ to quadratic order in $V$, we find $E_n = E_{n,0}+ E_{n,1} +  E_{n,2} + \ldots,$ with \cite{SI} 
\begin{equation}
     E_{n,1} = \langle n_0| V |n_0 \rangle \,\,\, ; \,\,\,  
     E_{n,2} =\sum_{m \neq n} \frac{|\matrixel{n_0}{V}{m_0}|^2}{2\tan \frac{E_{n,0}-E_{m,0}}{2}}. 
     \label{eq:pt}
\end{equation}
Here, we assume nondegenerate eigenstates $\ket{n_0}$ of $U_{F,0}$ with eigenphases $E_{n,0}$. For degenerate eigenstates, one first diagonalizes $V$ within the degenerate subspace. Importantly, coupling to a closeby level with small eigenphase difference
$\delta_0$ gives a small denominator in $E_{n,2}$. In contrast, coupling to a level with eigenphase difference  $\pi-\delta_\pi$ close to $\pi$ gives a large eigenphase denominator $\tan \frac{\pi-\delta_\pi}{2} \simeq \frac{2}{\delta_\pi}$. Indeed, the two states repel both ways around the unit circle [see Fig.\ \ref{fig1}(d)], suppressing the second-order correction and pushing the splitting closer to $\pi$. As we show below, $\pi$ pairings remain much more robust than  zero pairings for many-level systems. 

As the $Z_j$ are odd under the spin-flip (fermion-parity) symmetry $P$, a longitudinal field $V=\frac{\pi}{2}\sum_{j=1}^N h_j Z_j$ generically has a nonzero matrix element $\langle n^e_0|V|n^o_0\rangle$ coupling partner states, but zero diagonal matrix elements. Here, we denote the two paired many-body eigenstates of $U_{F,0}$ as $\ket{n_0^e}$ and $\ket{n_0^o}$. They have the same occupations $\gamma_\alpha^\dagger\gamma_\alpha^{\phantom{\dagger}}$ except for the Majorana occupation $n_{0,\pi}= \gamma_{0,\pi}^\dagger \gamma^{\phantom{\dagger}}_{0,\pi}$, with $n_{0,\pi}=0$ for $\ket{n^e_0}$ and $n_{0,\pi}=1$ for $\ket{n^o_0}$. 

In the MZM phase, we can restrict to the two paired levels provided that hybridization  splitting and perturbation are small compared to the level spacing of the many-body spectrum. Diagonalizing the Hamiltonian within this near-degenerate subspace gives the perturbed splitting
$\delta_0^\prime=\sqrt{\delta_0^2+4|\langle n^e_0|V|n^o_0\rangle|^2}$. This interpolates between second- 
and first-order perturbation theory as the random field $V$ increases. The eigenstates evolve into perturbed eigenstates $\ket{n_\pm} \simeq  \frac{1}{\sqrt{2}}(\ket{n_0^e}\pm \ket{n_0^o})$, once the perturbation exceeds $\delta_0$. With this understanding, we derive an analytical  splitting distribution \cite{SI}, which is in excellent agreement with numerical results [Fig.\ \ref{fig2}(c)]. Here, the square-root singularity of the splitting distribution at $\delta'_0=\delta_0$ is generic, while the bulk of the distribution is sensitive to the choice for the distribution of the random fields. 

\begin{figure*}[t!]
\centering
\includegraphics[width=\textwidth]{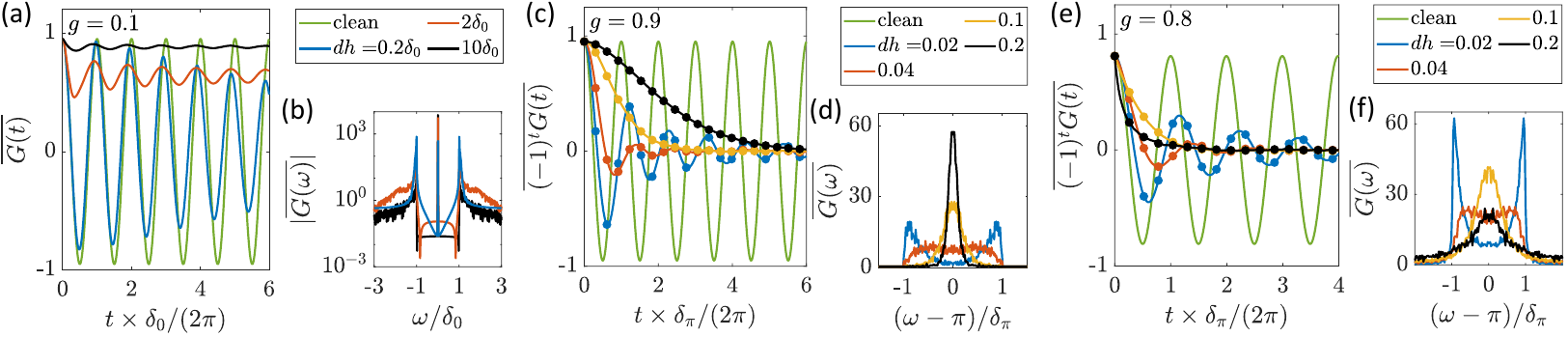}
      \caption{Boundary spin-spin correlation function $G(t)$ and its Fourier transform $G(\omega)$ with random longitudinal fields (see legends for strength). (a),(b) MZM phase: The random field suppresses the oscillations induced by the finite splitting $\delta_0$
      and generates a constant ($\omega=0$) contribution.  (c) MPM phase at $g=0.9$: The random field suppresses the  oscillations induced by $\delta_\pi$ in $(-1)^tG(t)$. The decay is Gaussian for large disorder and becomes slower with increasing $dh$. The correlation function (markers) is well reproduced when restricting the summation in Eq.\ (\ref{eq:ZZ}) to $\pi$-paired states $n$ and $m$, and (d) $ G(\omega)$ approximately tracks the $\delta_\pi^\prime$ distribution, cf.\ Fig.\ \ref{fig2}(d).   (e),(f) MPM phase at $g=0.8$: $(-1)^tG(t)$ now decays  exponentially reflecting the Lorentzian $\delta_\pi^\prime$ distribution. Parameters: $J=0.5$, $N=12$, $\mathcal{N}=10$, (a)-(d) $\delta_0=\delta_\pi=5\times 10^{-10}$, (e),(f) $\delta_\pi=2\times 10^{-6}$.} 
\label{fig3}
\end{figure*}

In the MPM phase, the coupling between the two $\pi$-paired states 
is negligible. Thus, we retain coupling between states belonging to different pairs. Evaluating the splittings $\delta_n=E_n^e-E_n^o+\pi$ in second-order perturbation theory, we find 
\begin{eqnarray}
    &&\delta_n \simeq \delta_\pi + 
    \sum_{m}\left\{
    \frac{|\matrixel{n^e_0}{V}{m_0^o}|^2}{2\tan \frac{E_n^e-E_m^o}{2}}-\frac{|\matrixel{n^o_0}{V}{m^e_0}|^2}{2\tan \frac{E_n^o-E_m^e}{2}}\right\}
    \nonumber\\
    &&\qquad
    +\sum_{m\neq n}\left\{
    \frac{|\matrixel{n^e_0}{V}{m_0^e}|^2}{2\tan \frac{E_n^e-E_m^e}{2}}-\frac{|\matrixel{n^o_0}{V}{m^o_0}|^2}{2\tan \frac{E_n^o-E_m^o}{2}}\right\}
    .
    \label{eq:2ndpt}
\end{eqnarray}
We have made second-order perturbation theory self-consistent by inserting the exact eigenphases $E_n^{e,o}$ in the denominators. This is motivated by the observation that there are couplings between many different pairs, which are of similar magnitude and can thus plausibly be accounted for in a self-consistent scheme. Linearizing Eq.\ (\ref{eq:2ndpt}) in the small splittings $\delta_n$, it can be readily solved numerically \cite{SI}. Figure \ref{fig2}(d) shows that the resulting splitting distribution  reproduces exact diagonalization data remarkably well over a wide range of disorder strengths. In particular, one reproduces the crossover from a bimodal distribution peaked near the splittings of the clean system to a narrower distribution peaked at $\delta'_\pi=0$ [Fig.\ \ref{fig2}(d),(e)] with increasing disorder $dh$. We observe that the distribution peaked at $\delta'_\pi=0$ is approximately Gaussian, when $g$ is sufficiently close to unity, but becomes Lorentzian for larger $1-g$. 

A corresponding ``phase diagram" is shown in Fig.\ \ref{fig2}(f), which can be understood from Eq.\ (\ref{eq:2ndpt}). For $g$ close to unity, the second sum on the right hand side can be dropped. Then, expanding in the $\delta_n$, we find $\delta_n=\delta_\pi - \sum_m (\delta_n+\delta_m) \Sigma_{nm}$, where the
\begin{equation} \Sigma_{nm} =  \frac{|\matrixel{n^e_0}{V}{m^o_0}|^2}{4\cos^2 \frac{E_n^e-E_m^e}{2}}
\label{eq:lambdapt}
\end{equation}
are strictly positive. Setting $\delta_n\approx \delta_\mathrm{typ}$ as well as $\delta_m\approx \pm\delta_\mathrm{typ}$
, the typical splitting
$\delta_\mathrm{typ}\approx\delta_\pi/(1+\langle\sum_m\Sigma_{nm}\rangle_n)$ is indeed reduced compared to $\delta_\pi$. ($\langle\ldots\rangle_n$ is an average over $n$.) The crossover between the bi- and unimodal distributions occurs when $\Sigma \sim 1$, implying  $N^{**}\propto \ln(1/dh)$, approximately independent of $g$ \cite{SI}. 

As $g$ deviates further from unity, the single-particle band broadens [Fig.\ \ref{fig1}(a)]. Consequently, the many-body eigenphases cover the entire interval $[-\pi,\pi]$ when $N > N^*\sim 1/(1-g)^2$. In this regime, the second term on the right hand side of Eq.\ (\ref{eq:2ndpt}) becomes significant due to the appearance of small denominators. The Lorentzian distribution can then be interpreted as an instance of a stable Levy distribution \cite{SI,Bouchaud1990}. We note that the splitting is well defined when the Majorana splitting $\sim e^{-N/\xi}$ is small compared to the many-body level spacing $\sim 2^{-N}$, where $\xi$ is the correlation length of the clean model, a condition satisfied for $g > 0.71$ at $J=0.5$. 

{\em Boundary spin-spin correlations.---}We finally consider the boundary spin-spin correlation function 
\begin{equation}
    G(t) = \frac{1}{2^N}\mathrm{tr} \{ Z_1(t)Z_1(0) \}
    =\frac{1}{2^N}\sum_{n,m}|(Z_1)_{nm}|^2 e^{-iE_{nm}t}
,\label{eq:ZZ}
\end{equation}
averaged over all initial states. Here, $(Z_1)_{nm}=\langle n | Z_1|m\rangle$ and $E_{nm}=E_{n}-E_{m}$. Sums are over all $2^N$ many-body eigenstates. In the MZM phase of the clean model, the pairing of eigenphases makes $G(t)$ oscillate with an exponentially long period $1/\delta_0$, [Fig.\ \ref{fig1}(e)]. In the MPM phase, the slow oscillations with period $1/\delta_\pi$ modulate rapid period-two oscillations [Fig.\ \ref{fig1}(f)]. 

Experimentally, $G(t)$ in the presence of a random longitudinal field persists up to times of the order of the qubit lifetime ($\ll 1/\delta_{0,\pi}$) regardless of the phase \cite{Mi2022}. This is surprising given the dramatically different sensitivities of the zero and $\pi$ pairings to longitudinal disorder. In fact, we find that the reasons underlying the robustness of $G(t)$ are very different in the two phases and that the long-time behaviors are actually quite distinct. 

In the MZM phase, the longitudinal field effectively polarizes the boundary spins. Spins located away from the boundary remain unpolarized due to the presence of mobile domain walls in generic states. Correspondingly, first-order degenerate perturbation theory gives perturbed eigenstates $\ket{n_\pm}$, which have nonzero diagonal matrix elements of $Z_1$ and $Z_N$. Then, the boundary spin-spin correlation function in Eq.\ (\ref{eq:ZZ}) has diagonal and thus time-independent terms, once the perturbation is large compared to the exponentially small splitting [Fig.\ \ref{fig3}(a)]. The Fourier transform of the boundary spin-spin correlation function develops a dominant zero-frequency peak [Fig.\ \ref{fig3}(b)]. In parallel, longitudinal disorder rapidly suppresses the amplitude of the Majorana oscillations. 

In the MPM phase, we observe that the period-two oscillations persist in the presence of a random longitudinal field, while the slow oscillation of their envelope decays, see Fig.\ \ref{fig3}(c),(e). This can be understood as a consequence of the splitting distribution across the many-body spectrum akin to inhomogeneous broadening. In fact, $G(t)$ in Eq.\ (\ref{eq:ZZ}) is dominated by terms, in which $\ket{n}$ and $\ket{m}$ are $\pi$-paired states [Fig.\ \ref{fig3}(c),(e)]. Then, the envelope of $G(t)$ is effectively the Fourier transform of the splitting distribution [Fig.\ \ref{fig3}(d),(f)]. Damped oscillations of the envelope persist for a bimodal distribution, with a long-time power-law tail due to the hard cutoff of the splitting distribution at $\delta^\prime_\pi=\delta_\pi$. This gives way to a nonoscillatory Gaussian (exponential) decay in the Gaussian (Lorentzian) regimes of the splitting distribution [Fig.\ \ref{fig2}(f)]. Thus, in the MPM phase, $G(t)$ directly reflects the robustness of the $\pi$ pairing to a random longitudinal field. 

\textit{Conclusions.---}We showed that even in the presence of random longitudinal fields far exceeding the nominal MPM splitting $\delta_\pi$, the $\pi$ pairing of the MPM phase remains exponentially precise in the system size $N$. We explain this surprising robustness, which contrasts sharply with the sensitivity of the zero pairing in the MZM phase, in terms of level repulsion of many-body Floquet levels on the unit circle, without invoking the notion of prethermalization \cite{Mi2022,Else2017,PhysRevB.95.014112}. 

It has been suggested to exploit the zero pairing in the quantum Ising model for realizing qubits, e.g., by implementing the model in chains of Josephson junctions \cite{Bruder1993,Levitov2001,You2014}. However, unlike the closely related Majorana qubits \cite{Plugge2017,Karzig2017,Oreg2020}, there would be no  protection against symmetry-breaking longitudinal fields. This may make the remarkable robustness of $\pi$ pairing interesting for applications in quantum information processing. 

\begin{acknowledgments}
We thank Piet Brouwer for an insightful discussion. Financial support was provided by Deutsche Forschungsgemeinschaft through CRC 183 and a joint ANR-DFG project (TWISTGRAPH) as well as the Einstein Research Unit on Quantum Devices at Freie Universit\"{a}t Berlin as well as by NSF Grant No.~DMR-2002275 
and the Army Research Office (ARO) under grant number W911NF-23-1-0051 at Yale University. L.I.G.\ thanks Freie Universit\"{a}t Berlin for hosting him as a CRC 183 Mercator fellow. We thank the HPC service of ZEDAT, Freie Universit\"{a}t Berlin, for computing time \cite{Bennett2020}. 
\end{acknowledgments}


%

\onecolumngrid

\clearpage

\setcounter{figure}{0}
\setcounter{section}{0}
\setcounter{equation}{0}
\renewcommand{\theequation}{S\arabic{equation}}
\renewcommand{\thefigure}{S\arabic{figure}}

\onecolumngrid

\section{Supplemental Material}

\section{Floquet quantum Ising model in the absence of random fields}

We review the mapping of the one-dimensional Floquet quantum Ising model to a Floquet Kitaev chain in the absence of disorder. The Floquet operator of the quantum Ising model [Eq.\ (\ref{eq:U_F0}) of the main text] obeys spin-flip symmetry (so that eigenstates can be classified into even and odd with respect to the spin-flip operator $P=\prod^N_{j=1}X_j$). 
For even $N$, it obeys a charge-conjugation symmetry (operator $C = i^{\frac{N}{2}} (\prod^{\frac{N}{2}}_{j=1}Y_{2j-1}Z_{2j})\mathcal{K}$ involving complex conjugation $\mathcal{K}$, implying that  eigenvalues come in complex-conjugate pairs). Moreover, the  isospectral symmetrized Floquet operator $U^s_{F,0}=U_{g/2}U_JU_{g/2}$ obeys a time-reversal symmetry (operator $\mathcal{K}$, implying that one can choose a real eigenbasis).

The Jordan-Wigner transformation 
\begin{align}
    \sigma_j^- = e^{i \pi \sum_{l<j} c^\dagger_lc_l}c_j,\qquad
    X_j = 1-2c^\dagger_j c_j,
    \label{eq:jwt}
\end{align}
with $\sigma_j^\pm = \frac{1}{2}(Z_j \pm i Y_j )$ maps the spin operators in Eq.\ \eqref{eq:U_F0} to  fermions $c_j$. The Floquet operator $U_{F,0}$ maps to the Floquet   Kitaev chain 
\begin{align}
\label{eq: Floquet Kitaev chain}
    U_{F,0}=e^{i\frac{\pi g}{2}  \sum\limits^N_{j=1} (1 - 2 c^\dagger_j c_j )}  e^{i\frac{\pi J}{2}\sum\limits^{N-1}_{j=1} (c_{j+1} + c^\dagger_{j+1})( c_{j} - c_{j}^\dagger) }.
\end{align}
In the fermionic formulation, the spin-flip symmetry translates to conservation of  fermion parity [operator $P=\prod^N_{j=1} \big(1-2c^\dagger_jc_j\big)$]. 

To work out the time evolution $c_j(t+1)=U_{F,0}^\dagger c_j(t) U_{F,0}$ of the fermion operators, one writes the fermion operators $c_j=\frac{1}{2}(a_{2j-1}+ia_{2j})$ in terms of Majorana operators $a_j$, so that
\begin{equation}
 U_{F,0} = \prod_{j=1}^N e^{\frac{\pi g}{2}  a_{2j-1} a_{2j} }\prod_{j=1}^{N-1} e^{\frac{\pi J}{2}  a_{2j} a_{2j+1} } = \prod_{j=1}^N \left(
 \cos \frac{\pi g}{2} +   a_{2j-1} a_{2j} \sin \frac{\pi g}{2}\right)\prod_{j=1}^{N-1} \left( \cos\frac{\pi J}{2} +   a_{2j} a_{2j+1} \sin \frac{\pi J}{2}\right). 
\end{equation}
We can find single-particle Bogoliubov operators satisfying \begin{equation}
    U_{F,0}^\dagger \gamma_\alpha U_{F,0} = e^{-i\epsilon_\alpha} \gamma_\alpha
\end{equation}
(see main text), where the operators $\gamma_\alpha$ are linear combinations of the $c_j$ and $c_j^\dagger$ and the $\epsilon_\alpha$ define the single-particle eigenphases. For periodic boundary conditions, we pass to the momentum representation $c_k=\frac{1}{\sqrt{N}}\sum_{j}e^{ikj}c_j$. Introducing the two-component operator $\phi_k = [c_k,c^\dagger_{-k}]^T$, the time evolution takes the form   
\begin{equation}     \phi_k(t+1)=U_{\mathrm{BdG}}\phi_k(t),
\end{equation}
with the Bogoliubov-de Gennes Floquet operator
\begin{align}
     U_{\mathrm{BdG}}(k)
    =
    \begin{pmatrix}
   e^{-i\pi g}\big(\cos(\pi J)+i\sin(\pi J)\cos k \big)
    &
    e^{-i\pi g}\sin(\pi J)\sin k
    \\
   -e^{i\pi g}\sin(\pi J)\sin k
    &
    e^{i\pi g}\big(\cos(\pi J)-i \sin(\pi J)\cos k\big)
    \end{pmatrix}.
    \label{eq: bloch-floquet matrix}
\end{align}
Diagonalizing $U_\mathrm{BdG} = D^\dagger \Lambda D$ with a diagonal matrix $\Lambda$, we have $\phi_k^\prime (t+1) = \Lambda \phi_k^\prime (t)$ with $\phi_k^\prime = D \phi_k$. We can thus identify the entries of $\phi_k^\prime = [\gamma_k,\gamma_{-k}^\dagger]^T$ with the Bogoliubov operators. An explicit calculation gives the particle-hole symmetric single-particle spectrum 
\begin{align}
\label{eq: single-particle spectrum}
\cos \epsilon_k = \cos(\pi J)\cos(\pi g)+\sin(\pi J)\sin(\pi g) \cos k, 
\end{align}
with the eigenphases $\epsilon_k$ defined modulo $2\pi$. 

Bulk gap closings signal phase transitions and  occur at  $\epsilon=0$ or $\epsilon=\pi$. Due to the invariance of $U_{F,0}$ under $g \rightarrow g+2$ and $J \rightarrow J+2$ as well as  $g\rightarrow -g$ and $J\rightarrow -J$, we can restrict attention to   $0\leq g,J\leq 1$. For these parameters, the spectral gap  $\Delta_0 = \pi(g - J)$ at zero energy is topological for $g<J$. Likewise, the spectral gap $\Delta_\pi= \pi(g+J-1)$ at $\pi$ is topological for $1-g>J$. This gives the phase diagram in Fig.\ \ref{fig1}(b) of the main text. 

In the fermion model with open boundary conditions, bulk gap closings indicate transitions between phases with and without localized Majorana modes at the ends. Majorana zero modes (MZMs) commute with the Floquet drive, while Majorana $\pi$ modes (MPMs) anticommute 
\begin{align}
    U^\dagger_{F,0}\gamma_0  U_{F,0} = \gamma_0, \qquad 
    U^\dagger_{F,0}\gamma_\pi  U_{F,0} = -\gamma_\pi.
\end{align}
The Majorana operators are odd under fermion parity $P$. Using the transfer-matrix technique, one can construct explicit Majorana operators for semi-infinite chains  \cite{Lerose2021}. In particular, one finds the localization lengths
\begin{equation}
 \xi_{0,\pi} = -\frac{1}{\ln{\lambda_{0,\pi}}}, \quad 
    \lambda_0 = \frac{\tan{\frac{\pi g}{2}}}{\tan{\frac{\pi J}{2}}}, \quad 
    \lambda_\pi = \frac{\cot{\frac{\pi g}{2}}}{\tan{\frac{\pi J}{2}}}
    \label{eq:xi0pi}
\end{equation}
of the MZM ($\xi_0$) and MPM ($\xi_\pi$) modes. In a finite chain, Majorana  hybridization leads to a splitting away from zero or $\pi$, which is exponentially small in the chain length, $\delta_{0,\pi} \propto e^{-N/\xi_{0,\pi}}$. One notices that the correlation lengths $\xi_0$ and $\xi_\pi$ map onto each other under $g \leftrightarrow 1-g$, explaining the symmetry of the Majorana splittings in Fig.\ \ref{fig1}(c) of the main text. We note that the Majorana splittings are smaller than the average many-body level spacing $2\pi/2^{N}$ provided that $\xi_{0,\pi}<1/\ln 2$. For $J=0.5$, this is true provided that $g>0.71$.

\section{Random transverse fields}

We find that the lognormal splitting distributions for a random transverse field as shown in Fig.\ \ref{fig2}(a),(b) are very well fit by 
\begin{equation}
\overline{\ln \frac{\delta_{0,\pi}}{\Delta}}=-\frac{N}{\xi_{0,\pi}},
    \qquad    \mathrm{var}\, \ln \frac{\delta_{0,\pi}}{\Delta}=\frac{N}{\ell}.   \label{eq:lognormal}
\end{equation}
Here, $\xi_{0,\pi}$ is the Majorana localization length in Eq.\ (\ref{eq:xi0pi}) and  
the mean free path $\ell$  can be accurately fit by 
\begin{equation}
     \ell=\frac{3}{\pi^2 (dg)^2}\sin^2(\pi g)
\end{equation}
across both phases. (Here, $dg$ denotes the width of the distribution of the random transverse field, see main text.) These results are closely analogous to results for the corresponding Hamiltonian problem  \cite{Brouwer2011}.

\section{Stroboscopic Floquet perturbation theory}

We derive the perturbative expressions given in Eq.\ (\ref{eq:pt}) in the main text. 
Guided by Hamiltonian perturbation theory, we expand eigenvalues and eigenstates of Floquet operators 
\begin{equation}
e^{-i\lambda V} U_0 \ket{n} = e^{-iE_n} \ket{n}
\label{eq:eiv}
\end{equation}
in powers of the perturbation $V$ as counted by powers of $\lambda$. Inserting the expansions 
\begin{equation}
    E_n = E_{n,0}+\lambda E_{n,1} + \lambda^2 E_{n,2} + \ldots,\qquad   \ket{n} = \ket{n_0} + \lambda \ket{n_1} + \lambda^2 \ket{n_2}+ \ldots 
\end{equation}
in Eq.\ (\ref{eq:eiv}), we find to quadratic order
\begin{align}
    &\bigg(1-i\lambda V - \frac{1}{2} \lambda^2 V^2 + \ldots \bigg)U_0  \big(\ket{n_0} + \lambda \ket{n_1} +  \lambda^2 \ket{n_2}  \ldots \big)
    \notag
    \\
    & \qquad = e^{-iE_{n,0}} \bigg(1-i\lambda E_{n,1} - \lambda^2 \big( i E_{n,2} + \frac{1}{2}E_{n,1}^2\big) + \ldots \bigg) \big(\ket{n_0} + \lambda \ket{n_1} +  \lambda^2 \ket{n_2}  \ldots \big).
\end{align}
We compare terms on both sides order by order in $\lambda$. At zeroth order, we recover  
\begin{equation}
    U_0 \ket{n_0} = e^{-i E_{n,0}} \ket{n_0}.
\end{equation}
At first order, we obtain (exploiting the orthogonality  $\braket{n_0}{n_1} = 0$) the first order shift
\begin{equation}
    E_{n,1} = \matrixel{n_0}{V}{n_0}
\end{equation}
as well as the first-order correction of the eigenstate,
\begin{equation}
    \ket{n_1} = ie^{-iE_{n,0}}\sum_{m \neq n} \frac{\matrixel{m_0}{V}{n_0}}{e^{-iE_{n,0}}-e^{-iE_{m,0}}}\ket{m_0}.
\end{equation}

The second-order correction to the eigenphases follows from  
\begin{equation}
    U_0\ket{n_2} - iVU_0\ket{n_1} -\frac{1}{2}V^2U_0 \ket{n_0} = e^{-iE_{n,0}}\left\{\ket{n_2} - iE_{n,1}\ket{n_1} -\left(iE_{n,2} +\frac{E_{n,1}^2}{2}\right)\ket{n_0}\right\}.
\end{equation}
Projecting this expression on  $\bra{n_0}$ and using $\langle n_0|n_2\rangle=0$ as well as the lower-order results, this yields 
\begin{equation}
    E_{n,2} = \sum_{m \neq n} \frac{|\matrixel{n_0}{V}{m_0}|^2}{2\tan \frac{E_{n,0}-E_{m,0}}{2}}.
    \label{eq:2ndopt}
\end{equation}
This reduces to the standard expressions of Hamiltonian perturbation theory for a pair of close levels. However, it differs  drastically from Hamiltonian perturbation theory when the difference between the unperturbed eigenphases is close to $\pi$, where the eigenphase denominator diverges. More generally the tangent accounts for the periodic nature of the eigenphases. 

\section{Splittings of paired many-body states}

Here, we provide more details on
Eqs.\ (\ref{eq:2ndpt}) and (\ref{eq:lambdapt}), which apply stroboscopic Floquet perturbation theory to the splittings of paired many-body eigenstates with $U_{0}= U_g U_J$ and $e^{iV}=U_h$. In the absence of a random longitudinal field, the paired states differ in their occupations of the fermion mode constructed from the Majorana (zero or $\pi$) modes. Thus, the (even and odd) states differ in the corresponding Majorana parity, 
\begin{align}
    \left(-i\gamma_L\gamma_R\right)\ket{n^\mathrm{e}}=\ket{n^\mathrm{e}}, \quad \left(-i\gamma_L\gamma_R\right)\ket{n^\mathrm{o}}=-\ket{n^\mathrm{o}},
\end{align}
with $\gamma_{L/R}$ denoting the Majorana operators at the left and right ends. Paired states convert into each other by application of the Majorana operators, e.g., $\gamma_L \ket{n^\mathrm{e}}=\ket{n^\mathrm{o}}$ and have identical occupations of all non-Majorana modes. The energies   $E\even_{n}$ and   $E\odd_{n}$ of the paired states differ by 
\begin{align}
     E\odd_{n}=E\even_{n}-\delta_0 \quad\mathrm{or} \quad E\odd_{n}=E\even_{n}+\pi-\delta_\pi,
\end{align}
in the case of MZM or MPM phases, respectively. The splittings $\delta_{0,\pi}$ are exponentially small in the length of the chain, $\delta_{0,\pi}\sim e^{-N/\xi_{0,\pi}}$ and identical for all pairs. 

The random longitudinal field 
\begin{align}
V=\sum_j h_j Z_j,
\end{align}
is odd under total fermion parity. It thus couples unperturbed states, which have  different total parity $P$, but may have identical or different occupations of the Majorana mode. 

\subsection{Splittings of MZM modes}

In discussing the perturbed  splittings of MZM modes, we assume that the field is much smaller than the many-body level spacing. Then we can restrict attention to a pair of partner states. The coupling between the states is dominated by the effects of the fields $h_1$ and $h_N$ acting on the spins at the ends of the chain. While the boundary spins $Z_1$ and $Z_N$ are polarized by the random longitudinal field, the interior spins remain unpolarized due to the existence of mobile domain walls in generic eigenstates. Technically, this suppression arises from the string operators in Eq.\ (\ref{eq:jwt}). Thus, we have 
\begin{align}
    v=\mel{n\even}{V}{n\odd}\approx \frac{\pi h_1}{2} \mel{n\even}{Z_1}{n\odd}+\frac{\pi h_N}{2}\mel{n\even}{Z_N}{n\odd}=\frac{\pi(h_1+h_N) \psi_M }{2}
\end{align}
for the matrix elements entering the effective $2\times 2$ Hamiltonian. Here, we used that $Z_1 = \psi_M \gamma_L + \dots$ and $Z_N = i \psi_M \gamma_R P + \dots$, 
where $\psi_M$ is the Majorana wavefunction at the boundary sites, $P$ denotes the fermion parity operator, and the ellipses stand for above-gap excitations. For uniformly distributed $h_1,h_N \in [-dh,dh]$,  the matrix element  $v$ has a triangular distribution 
\begin{align}
    p(v)=\frac{1}{v_0}\left(1-\frac{\abs{v}}{v_0}\right)\theta\left(v_0-\abs{v}\right). 
\end{align}
Here, $v_0=2\psi_M (\pi dh /2)$ is the maximal shift caused by the two boundary fields and $\theta(x)$ is the Heaviside function. The effective $2\times 2$ Hamiltonian becomes
\begin{align}
    H_n^\mathrm{eo}=\frac{\delta_0}{2}\bigg(\ket{n\even}\bra{n\even}-\ket{n\odd}\bra{n\odd}\bigg)+v\bigg(\ket{n\even}\bra{n\odd}+\ket{n\odd}\bra{n\even}\bigg),
\end{align}
where $\delta_0>0$ is the bare splitting. Note that the Hamiltonian takes the same form for all pairs $n$, so that the splitting remains uniform across the many-body spectrum (to leading order) and varies only between disorder realizations. The eigenenergies  $E_\pm=\pm \frac{1}{2}\sqrt{\delta_0^2+4v^2}$, yield the perturbed splittings
\begin{align}
    \delta^\prime_0=E_+-E_-=\sqrt{\delta_0^2+4v^2}, 
\end{align}
which are  larger than the bare splittings. Using $p(v)$, we arrive at
\begin{align}
p\left(\delta^\prime_0\right)=\frac{\abs{\delta_0^\prime}}{2v_0}\left(\frac{1}{\sqrt{\delta_0^\prime\phantom{}^2-\delta_0^2}}-\frac{1}{2v_0}\right)   \theta\left(\abs{\delta_0^\prime\phantom{}}-\delta_0\right)\theta\left(\sqrt{\delta_0^2+4v_0^2}-\abs{\delta_0^\prime}\right).
\end{align}
At $\delta^\prime_0 = \delta_0$ we find a square-root singularity, which -- unlike the bulk of the distribution -- is insensitive to the specific choice of distribution of the random fields. 

\subsection{Splittings of MPM modes}
 
In the MPM phase, the Majorana modes do not induce  degeneracies in the many-body spectrum of $U_{F,0}$. Hence, the random longitudinal field affects the spectrum only in second-order perturbation theory. We write
\begin{align}
     E\odd_{n}=E\even_{n}+\pi-\delta_n. 
\end{align}
The random field shifts the splitting $\delta_n$ away from the bare splitting $\delta_\pi$ by an amount $\Delta \delta_n$, 
\begin{align}
    \delta_n=\delta_\pi +\Delta \delta_n. 
\end{align}
In second-order perturbation theory, Eq.\ (\ref{eq:2ndopt}), the shift becomes 
    \begin{align}
\label{eq: splitting general}
    \Delta \delta_{n} = \sum_{m}
    \left[\frac{ \abs{v^\mathrm{eo}_{nm}}^2}
    {2\tan \dfrac{E^\mathrm{e}_n-E^\mathrm{o}_m}{2} }
    -
    \frac{\abs{v^\mathrm{oe}_{nm}}^2}
    {2\tan \dfrac{E^\mathrm{o}_n-E^\mathrm{e}_m}{2} }\right]
    +
    \sum_{m\neq n}
   \left[\frac{ \abs{v^\mathrm{ee}_{nm}}^2}{2\tan \dfrac{E^\mathrm{e}_n-E^\mathrm{e}_m}{2} }-\frac{\abs{v^\mathrm{oo}_{nm}}^2}{2\tan \dfrac{E^\mathrm{o}_n-E^\mathrm{o}_m}{2} }\right]
\end{align} 
with matrix elements
\begin{align}
    v^{ab}_{nm} = \mel{n^a}{V}{m^b}, \quad a,b\in \{ \mathrm{e} ,\mathrm{o}\}.
\end{align}
The first term contains processes which change  the Majorana parity $-i\gamma_L\gamma_R$, in  addition to the global fermion parity $P$. Consequently, the bulk parity defined as  $Q=(-i\gamma_L\gamma_R)P$ remains invariant. The second term contains processes which leave the Majorana parity $-i\gamma_L\gamma_R$ unchanged, implying that 
$Q$ changes. 

In the MPM phase, the coupling within the pairs is negligible due to the divergence of the eigenphase denominator as the eigenphase difference approaches $\pi$. Thus, the effect of  the perturbation is controlled by the coupling between different pairs. Since there are many such couplings of similar magnitude, it is plausible that their effect can be approximated in a self-consistent scheme. For this reason, we made the perturbative expression 
in Eq.\ (\ref{eq: splitting general}) self-consistent (in analogy with the self-consistent Born approximation) by retaining the exact eigenenergies $E_{n}^{e/o}$ in the denominators. 

Using that the splittings $\delta_n$ are small, we expand the right-hand side of Eq.\ (\ref{eq: splitting general}) for $\Delta\delta_n$ to linear order in the $\delta_n$. This yields
\begin{equation}
    \delta_n -
    \delta_\pi = - \sum_m\Sigma_{nm}\delta_m + \Lambda_n, 
    \label{eq:deltadelta1}
\end{equation}
with
\begin{equation}
\Sigma_{nm} =   \left( \sum_l   \frac{\abs{v^\mathrm{eo}_{nl}}^2}{4\cos^2 \dfrac{E\even_n-E\even_l}{2}}
    -
    \sum_{l\neq n}\frac{\abs{v^\mathrm{oo}_{nm}}^2}{4\sin^2 \dfrac{E\even_n-E\even_l}{2}}\right)\delta_{nm}
        +    \frac{\abs{v^\mathrm{eo}_{nm}}^2}{4\cos^2 \dfrac{E\even_n-E\even_m}{2}}
    +
    \frac{\abs{v^\mathrm{oo}_{nm}}^2(1-\delta_{nm})}{4\sin^2 \dfrac{E\even_n-E\even_m}{2}}
    \label{eq:sigmanm}
\end{equation}
and
\begin{equation}
\Lambda_n = \sum_{m\neq n}
   \frac{ \abs{v^\mathrm{ee}_{nm}}^2-\abs{v^\mathrm{oo}_{nm}}^2}
    {2\tan \dfrac{E\even_n-E\even_m}{2} }.
    \label{eq:deltanm}
\end{equation}
We can then express the vector $\delta^\prime_{\pi}$ of splittings $\delta_n$ in matrix notation as 
\begin{equation}
    \delta^\prime_{\pi} = \frac{1}{1 + \Sigma} (\delta_\pi + \Lambda) \label{eq;matrix}
\end{equation}
Here, $\delta_\pi$ should also be interpreted as a vector, with all entries equal to the bare splitting $\delta_\pi$.

To derive these expressions, it is convenient to decompose the matrix elements into symmetric and antisymmetric matrix contributions,
\begin{equation}
    \label{eq: delta  sym,asym}
    \Delta \delta_{n} = \Delta \delta^{\mathrm{eo/oe}}_{n}+ \Delta \delta^{\mathrm{ee/oo}}_{n}
\end{equation}
with 
\begin{eqnarray}
    \Delta \delta^{\mathrm{eo/oe}}_{n} &=& \sum_{m}\frac{\abs{v^\mathrm{eo}_{nm}}^2+\abs{v^\mathrm{oe}_{nm}}^2}{2}
    \left[\frac{ 1}
    {2\tan \dfrac{E^\mathrm{e}_n-E^\mathrm{o}_m}{2} }
    -
    \frac{1}
    {2\tan \dfrac{E^\mathrm{o}_n-E^\mathrm{e}_m}{2} }\right]
    \nonumber\\
    && \qquad \qquad +\sum_{m}\frac{\abs{v^\mathrm{eo}_{nm}}^2-\abs{v^\mathrm{oe}_{nm}}^2}{2}
    \left[\frac{ 1}
    {2\tan \dfrac{E^\mathrm{e}_n-E^\mathrm{o}_m}{2} }
    +
    \frac{1}
    {2\tan \dfrac{E^\mathrm{o}_n-E^\mathrm{e}_m}{2} }\right]
\end{eqnarray}
and     
\begin{eqnarray}
 \Delta \delta^{\mathrm{ee/oo}}_{n} &=& \sum_{m\neq n}\frac{ \abs{v^\mathrm{ee}_{nm}}^2+\abs{v^\mathrm{oo}_{nm}}^2}{2}
   \left[\frac{1}{2\tan \dfrac{E^\mathrm{e}_n-E^\mathrm{e}_m}{2} }-\frac{1}{2\tan \dfrac{E^\mathrm{o}_n-E^\mathrm{o}_m}{2} }\right]
   \nonumber\\
   &&\qquad\qquad +
    \sum_{m\neq n}\frac{ \abs{v^\mathrm{ee}_{nm}}^2-\abs{v^\mathrm{oo}_{nm}}^2}{2}   \left[\frac{1}{2\tan \dfrac{E^\mathrm{e}_n-E^\mathrm{e}_m}{2} }+\frac{1}{2\tan \dfrac{E^\mathrm{o}_n-E^\mathrm{o}_m}{2} }\right].
\end{eqnarray} 
The eigenphase differences in the denominators can be written as 
\begin{align}
E\odd_n-E\odd_m=E\even_n-E\even_m-\delta_n+\delta_m, \qquad 
  E\even_n-E\odd_m =  E\even_n-E\even_m-\pi+\delta_m.
\end{align}
Using the expansions \begin{align}    &\frac{1}{\tan(\frac{x-\pi+\delta_m}{2})}-\frac{1}{\tan(\frac{x+\pi-\delta_n}{2})}\simeq -\frac{(\delta_n+\delta_m)}{2\cos^2(\frac{x}{2})},
    \\
     &\frac{1}{\tan(\frac{x-\pi+\delta_m}{2})}+\frac{1}{\tan(\frac{x+\pi-\delta_n}{2})}\simeq \frac{2}{\tan(\frac{x-\pi}{2})} -\frac{(\delta_n-\delta_m)}{2\cos^2(\frac{x}{2})}
\end{align}
to linear order, one finds 
\begin{align}
    \Delta \delta^{\mathrm{eo/oe}}_{n}=  \sum_{m}
   \frac{ \abs{v^\mathrm{eo}_{nm}}^2-\abs{v^\mathrm{oe}_{nm}}^2}
    {2\tan \dfrac{E\even_n-E\even_m-\pi}{2} }
    -\sum_{m}\frac{(\delta_n+\delta_m)\left(\abs{v^\mathrm{eo}_{nm}}^2+\abs{v^\mathrm{oe}_{nm}}^2\right)}{8\cos^2 \dfrac{E\even_n-E\even_m}{2}}
    -\sum_{m}\frac{(\delta_n-\delta_m)\left(\abs{v^\mathrm{eo}_{nm}}^2-\abs{v^\mathrm{oe}_{nm}}^2\right)}{8\cos^2 \dfrac{E\even_n-E\even_m}{2}}
\end{align}
as well as
\begin{align}
    \Delta \delta^{\mathrm{ee/oo}}_{n}= \sum_{m\neq n}
   \frac{ \abs{v^\mathrm{ee}_{nm}}^2-\abs{v^\mathrm{oo}_{nm}}^2}
    {2\tan \dfrac{E\even_n-E\even_m}{2} }
    -\sum_{m\neq n}\frac{(\delta_n-\delta_m)\big(\abs{v^\mathrm{ee}_{nm}}^2+\abs{v^\mathrm{oo}_{nm}}^2\big)}{8\sin^2 \dfrac{E\even_n-E\even_m}{2}}
     +\sum_{m\neq n}\frac{(\delta_n-\delta_m)\big(\abs{v^\mathrm{ee}_{nm}}^2-\abs{v^\mathrm{oo}_{nm}}^2\big)}{8\sin^2 \dfrac{E\even_n-E\even_m}{2}}.
\end{align}
Unlike in the MZM case, the splittings in the MPM phase vary across the many-body spectrum, so that terms involving $\delta_n-\delta_m$ do not vanish. Collecting terms and using
$\abs{v^\mathrm{eo}_{nm}}=\abs{v^\mathrm{oe}_{nm}}$, we find 
\begin{align}
\label{eq: self-cons eq}
    \delta_n - \delta_\pi = - 
    \sum_{m}\frac{\big(\delta_n+\delta_m)\abs{v^\mathrm{eo}_{nm}}^2}{4\cos^2 \dfrac{E\even_n-E\even_m}{2}}
    -
    \sum_{m\neq n}\frac{\big(\delta_n-\delta_m\big)\abs{v^\mathrm{oo}_{nm}}^2}{4\sin^2 \dfrac{E\even_n-E\even_m}{2}}
     +
       \sum_{m\neq n}
   \frac{ \abs{v^\mathrm{ee}_{nm}}^2-\abs{v^\mathrm{oo}_{nm}}^2}
    {2\tan \dfrac{E\even_n-E\even_m}{2} }
\end{align}
and thus Eqs.\ (\ref{eq:deltadelta1}), (\ref{eq:sigmanm}) and (\ref{eq:deltanm}). Note that the first term on the right-hand side involves matrix elements between states of different Majorana parities, while the second and third terms involve matrix elements between states of equal Majorana parities. 

\subsection{Implications}

We find that terms involving matrix elements between states of equal Majorana parities can be neglected for $N<N^*(g)$. In this regime, the eigenphase differences in the denominators of the corresponding terms in Eqs.\ (\ref{eq:sigmanm}) and (\ref{eq:deltanm}) remain large compared to the many-body level spacing. In fact, coupled states must have different bulk parities. For $g$ close to unity, the eigenphase regions supporting  states with different bulk parities do not overlap, so that the denominators remain large. This is a consequence of the small bandwidth $\propto (1-g)$ of the singe-particle excitations about the phase $\pm \pi/2$, see Fig.\ \ref{fig1}(a) of the main text. For zero single-particle bandwidth, the eigenphases of states with different bulk parities differ by an odd multiple of $\pi/2$. A finite single-particle bandwidth changes the many-body eigenphases by an amount of order $\propto N^{1/2}(1-g)$ (originating from summing over $N$ terms with random signs). As long as this  change remains small compared to unity, there are no small denominators in the expression for $\Lambda_n$. Thus, we conclude that $N^*\propto 1/(1-g)^2$. 

When $N<N^*$, Eq.\ (\ref{eq: self-cons eq}) simplifies to 
\begin{align}
    \delta_n - \delta_\pi = - 
    \sum_{m}\frac{\big(\delta_n+\delta_m)\abs{v^\mathrm{eo}_{nm}}^2}{4\cos^2 \dfrac{E\even_n-E\even_m}{2}},
    \label{eq:deltadelta}
\end{align}
which corresponds to Eq.\ (\ref{eq:lambdapt}) of the main text. In the perturbative limit (bimodal regime), the $\delta_n$ remain close to $\delta_\pi$ and we find that the random longitudinal field reduces the splittings $\delta_n$ below  $\delta_\pi$. More generally, we can rewrite Eq.\ (\ref{eq:deltadelta}) as 
\begin{equation}
    \sum_m\sigma_{nm} \delta_n = \delta_\pi
\end{equation}
with 
\begin{equation}
    \sigma_{nm} = \left( 1 + \sum_l \frac{\abs{v^\mathrm{eo}_{nm}}^2}{4\cos^2 \dfrac{E\even_n-E\even_l}{2}} \right)\delta_{nm} + \frac{\abs{v^\mathrm{eo}_{nm}}^2}{4\cos^2 \dfrac{E\even_n-E\even_m}{2}}.
\end{equation}
We then find
\begin{equation}
    \delta_n = \sum_m (\sigma^{-1})_{nm} \delta_\pi. 
    \label{eq:MPMGauss}
\end{equation}
When the perturbation becomes sufficiently large (Gaussian regime), $\sigma$ is a ``random" matrix far from the unit matrix with exclusively nonnegative entries. Then, one expects the matrix elements $(\sigma^{-1})_{nm}$ of the inverse matrix to have ``random" signs. The approximately Gaussian distribution which we find numerically
[see, e.g., the curve for $dh=0.1$ in Fig.\ 2(d)], can then be roughly interpreted as a consequence of the central limit theorem.  We note that the matrix elements of $\sigma$ have a rather broad distribution as a consequence of near degeneracies of the eigenphase denominators. At the same time, the distribution of the matrix elements of $\sigma^{-1}$ do not have long tails. However, the matrix elements of $\sigma^{-1}$ are still rather structured. As a result, the central-limit argument is less accurate for a particular disorder realization, but applies with reasonable accuracy after averaging over disorder configurations. 
The transition between the bimodal and Gaussian regimes occurs when the $\sigma_{nm}$ become of order unity. We find numerically that $\sigma_{nm}$ is of order $dh^2 \exp(N/ \zeta)$ with $\zeta \approx 1.65$, which depends only weakly on $g$. Thus, the transition occurs at $N^{**} \sim \ln(1/dh^2)$.  

Conversely, for $N>N^*$, all terms in Eq.\ (\ref{eq: self-cons eq}) have to be retained when computing the splitting $\delta_n$. In this regime, terms involving matrix elements between states of equal Majorana parities can be viewed as a sum over many terms of the form $1/x$ (with $x$ representing the eigenphase denominators), where $x$ has a distribution that remains nonzero for $x=0$. Assuming that the terms are statistically independent, one then obtains a Lorentzian distribution for $\Lambda_n$. This follows since the distribution of $1/x$ has a long tail, with the Lorentzian being the relevant Levy stable distribution \cite{Bouchaud1990}. While we observe  deviations from Lorentzian behavior for $\Lambda_n$, we find that the distribution of $\delta_n$ can be well fit by a Lorentzian. Possibly, the distribution of $\delta_n$ is less influenced by the lognormal distribution of matrix elements of the random field, as the matrix elements appear both in the numerator (via $\Lambda_n$) and the denominator (via $\Sigma_{nm}$).  

\end{document}